\documentclass[prl,aps,floats,twocolumn,showpacs,
preprintnumbers,showkeys]{revtex4-1}

\usepackage[latin1]{inputenc}                    
\usepackage{graphicx}                            
\usepackage{latexsym}                            
\usepackage{amsfonts}                            
\usepackage{amssymb}                             
\usepackage{amsmath}                             
\usepackage[mathscr]{eucal}                      
\usepackage{dcolumn}                             
\usepackage{theorem}                             
\newcommand{\bc}{\begin{center}}
\newcommand{\ec}{\end{center}}
\newcommand{\be}{\begin{equation}}
\newcommand{\ee}{\end{equation}}
\newcommand{\bea}{\begin{eqnarray}}
\newcommand{\eea}{\end{eqnarray}}

\newcommand{\del}{\partial}

\begin{document}

\preprint{
\vbox{
\hbox{ADP-12-23/T790}
}}

\title[Sigma terms from an SU(3) chiral extrapolation]{Sigma terms from an SU(3) chiral extrapolation}
\author{P.E.~Shanahan}\affiliation{ARC Centre of Excellence in Particle Physics at the Terascale and CSSM, School of Chemistry and Physics, University of Adelaide,
  Adelaide SA 5005, Australia}
\author{A.W.~Thomas}\affiliation{ARC Centre of Excellence in Particle Physics at the Terascale and CSSM, School of Chemistry and Physics, University of Adelaide,
  Adelaide SA 5005, Australia}
\author{R.D.~Young}\affiliation{ARC Centre of Excellence in Particle Physics at the Terascale and CSSM, School of Chemistry and Physics,
  University of Adelaide, Adelaide SA 5005, Australia}

\begin{abstract}
  We report a new analysis of lattice simulation results for
  octet baryon masses in 2$+$1-flavor QCD, with an emphasis on a
  precise determination of the strangeness nucleon sigma term. A
  controlled chiral extrapolation of a recent PACS-CS Collaboration
  data set yields baryon masses which exhibit remarkable agreement
  both with experimental values at the physical point and with the
  results of independent lattice QCD simulations at unphysical meson
  masses. Using the Feynman-Hellmann relation, we evaluate sigma
  commutators for all octet baryons. The small statistical uncertainty, and considerably smaller
  model-dependence, allows a significantly more precise determination
  of the pion-nucleon sigma commutator and the strangeness sigma term
  than hitherto possible, namely $\sigma_{\pi
    N}=45 \pm 6$~MeV and $\sigma_s = 21 \pm 6$~MeV at the physical
  point.

\end{abstract}



\maketitle


The light-quark sigma terms provide critical information concerning the nature of explicit chiral symmetry breaking in QCD, as well as the decomposition of the mass of the nucleon~\cite{Ji:1994av}.  While these
physical observables are difficult to measure with conventional
probes, an accurate knowledge of the sigma terms is of essential
importance in the interpretation of experimental searches for dark matter
\cite{Bottino:1999ei,Ellis:2008hf,Giedt:2009mr,Underwood2012,Hill:2011be}.
Dark matter candidates, such as the favoured
neutralino, a weakly interacting fermion with mass of order 100~GeV or more, have interactions with hadronic matter which are essentially
determined by couplings to the light and strange quark sigma
commutators.

%

Experimentally, $\sigma_{\pi N}$ is determined from $\pi N$ scattering
through a dispersion relation analysis. Traditionally, the strange scalar form factor has
then been evaluated indirectly using $\sigma_{\pi N}$ and a best-estimate for the non-singlet
contribution $\sigma_0 = m_l\langle N |
\overline{u}u+\overline{d}d-2\overline{s}s |N \rangle$. These traditional evaluations have yielded a value for $\sigma_s$ as large as 300~MeV,
compared to 50~MeV for the light quark commutator, indicating that
as much as one third of the nucleon mass might be attributed to non-valence
quarks. This suggestion appears to be incompatible with widely-used constituent
quark models, and has generated much theoretical
interest over the last two decades.

The traditional method of determination of $\sigma_s$ is severely limited because it involves the small difference between $\sigma_{\pi N}$ (with its uncertainty) and $\sigma_0$ which is usually deduced in terms of $SU(3)$ symmetry breaking.
Even given a perfect determination of $\sigma_{\pi N}$, $\sigma_s$ will have an uncertainty of order $\sim 90$~MeV~\cite{Young:2009ps}. 
For that reason $\sigma_s$ has been
considered notoriously difficult to pin down.
In recent years, the best value for
$\sigma_s$ has seen an enormous revision. Advances in lattice QCD have
revealed a strange sigma term of
20-50~MeV~\cite{Young:2009zb,Ohki2009,Toussaint2009,Dinter:2012tt,Bali:2011ks,Dinter:2011zz,Horsley:2011wr,Durr:2011mp,Babich:2010at,Takeda:2010cw}, an order of
magnitude smaller than was previously believed.

In this Letter we use the finite-range regularisation (FRR) technique
to effectively resum the chiral perturbation
theory expansion of the quark mass dependence of octet
baryons. Fitting the resulting functions to recent lattice data, we extract
the scalar form factors by simple differentiation using the
Feynman-Hellmann theorem. Our technique allows comparison with recent
direct lattice QCD calculations of the flavor-singlet matrix elements
at unphysical meson masses~\cite{Bali2011}, with consistent
results. We report values of $\sigma_{\pi N}=45 \pm 6$~MeV and
$\sigma_s = 21 \pm 6$~MeV at the physical point.

The sigma terms of a baryon $B$ are defined as scalar form factors,
evaluated in the limit of vanishing momentum transfer. For each quark
flavor $q$,
\begin{equation}
\sigma_{Bq} = m_q \langle B | \overline{q} q | B \rangle; \hspace{4mm}
\overline{\sigma}_{Bq} = \sigma_{Bq}/M_B.
\label{eq:dimlesssigma}
\end{equation}
For the nucleon, the so-called $\pi N$ sigma commutator and the
strange sigma commutator are defined by
\begin{align}
\sigma_{\pi N} & = m_l \langle N | \overline{u}u + \overline{d}d | N \rangle, \\
\sigma_s & = m_s \langle N | \overline{s}s | N \rangle,
\end{align}
where $m_l = (m_u+m_d)/2$.


Following the technique described in
Refs.~\cite{Young:2009zb,Shanahan:2011}, we fit octet baryon mass data
recently published by the PACS-CS Collaboration~\cite{Aoki:2008sm}
using a chiral expansion:
\begin{equation}
M_B=M^{(0)}+\delta M_B^{(1)}+ \delta M_B^{(3/2)}+ \ldots
\end{equation}
Here, $M^{(0)}$ denotes the degenerate mass of the baryon octet in the
SU(3) chiral limit, $\delta M_B^{(1)}$ gives the correction linear in
the quark masses, and $\delta M_B^{(3/2)}$ represents quantum
corrections corresponding to one-loop contributions from the
pseudo-Goldstone bosons $\phi=\pi,K,\eta$. Explicit expressions for
the extrapolation formulae and for renormalisation of the loop
integrals may be found in Ref.~\cite{Shanahan:2011}.

Following Ref.~\cite{Young:2009zb}, we retain the octet-decuplet mass
difference $\delta$ in numerical evaluations to properly account for
the branch structure near $m_\phi \sim \delta$.  The loop contribution
parameters are set to appropriate experimental and phenomenological
values; $D+F=g_A=1.27$, $F=\frac{2}{3}D$, $C=-2D$, $f=0.0871$~GeV, and
$\delta=0.292$~GeV. Within the framework of FRR, we introduce a mass
scale $\Lambda$, through a regulator $u(k)$. $\Lambda$ is related to
the scale beyond which a formal expansion in powers of the Goldstone
boson masses breaks down. In practice, $\Lambda$ is chosen by fitting
to the lattice data itself. For further discussions of the FRR
regularization scheme, we refer to
Refs.~\cite{Leinweber:2003dg,Hall2011,Stuckey:1996qr,Donoghue:1998bs,Leinweber:1998ej}.
To provide an estimate of the model-dependent uncertainty in our
result, we consider a variety of forms of the regulator $u(k)$, namely
monopole, dipole, and Gaussian, as well as a sharp cutoff. To further estimate systematic uncertainties, we allow $f$, the meson decay constant in the chiral limit, the baryon-baryon-meson coupling constants $F$ and $C$, and
$\delta$ to vary by $\pm 10 \%$ from the central values given
above; see Ref.~\cite{Thomas2011} for details. The effect of these variations are included in the final quoted errors.

The PACS-CS results have been corrected for small, model-independent, finite
volume effects before fitting. These finite volume corrections were
evaluated by considering the leading one-loop results of chiral EFT~\cite{Young:2009zb,Geng2011,Beane:2011pc,Khan2004}. We note that the largest shift was $-0.022 \pm
0.002$~GeV for the nucleon at the lightest pion mass.

\begin{figure}[tbf]
\bc
\includegraphics[width=0.48\textwidth]{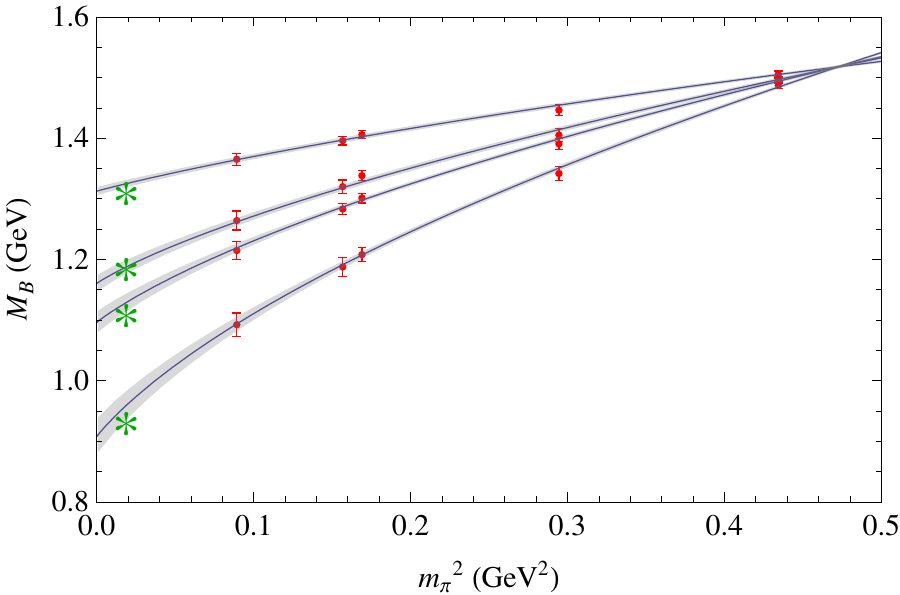}
\caption{Fit to the PACS-CS baryon octet data. Error
  bands shown are purely statistical, and incorporate correlated
  uncertainties between all fit parameters. Note that the data shown has been corrected for finite volume and the simulation strange
  quark mass, which was somewhat larger than the physical value. The green stars show
  experimental values.}
\label{fig:oct}
\ec
\end{figure}

The fit to the PACS-CS baryon octet data is shown in
Figure~\ref{fig:oct}. We find an optimal dipole regularization scale
of $\Lambda=0.9 \pm 0.1$~GeV, in close agreement with the value
deduced from an analysis of nucleon magnetic moment
data~\cite{Hall2012} and, from the phenomenological point of view,
remarkably close to the value preferred from comparison of the
nucleon's axial and induced pseudoscalar form factors~\cite{Guichon1983}. The minimum $\chi^2_{\textrm{dof}}$ is 0.41 (6.1/(20-5)) for the dipole, and varies between 0.40 and 0.42 for the
other regulators. This value is
somewhat lower than unity, as correlations between the lattice data
cannot be accounted for without access to the original data.

Clearly, the fit is very satisfactory over the entire
range of quark masses explored in the simulations. Furthermore, the
masses of the octet baryons agree remarkably with experiment at the
physical point. A comparison of the extrapolated baryon masses with
the best experimental values is given in Table~\ref{table:sigmas}. The
first error quoted is statistical and includes the correlated
uncertainty of all of the fit parameters including the regulator mass
$\Lambda$, while the second is an estimate of model-dependence. This
includes the full variation over dipole, monopole, sharp cutoff and
Gaussian regulator forms, as well as accounting for the
variation of the phenomenologically-set parameters $F$, $C$ and $\delta$ described earlier.

\begin{figure}[tbf]
\bc
\includegraphics[width=0.48\textwidth]{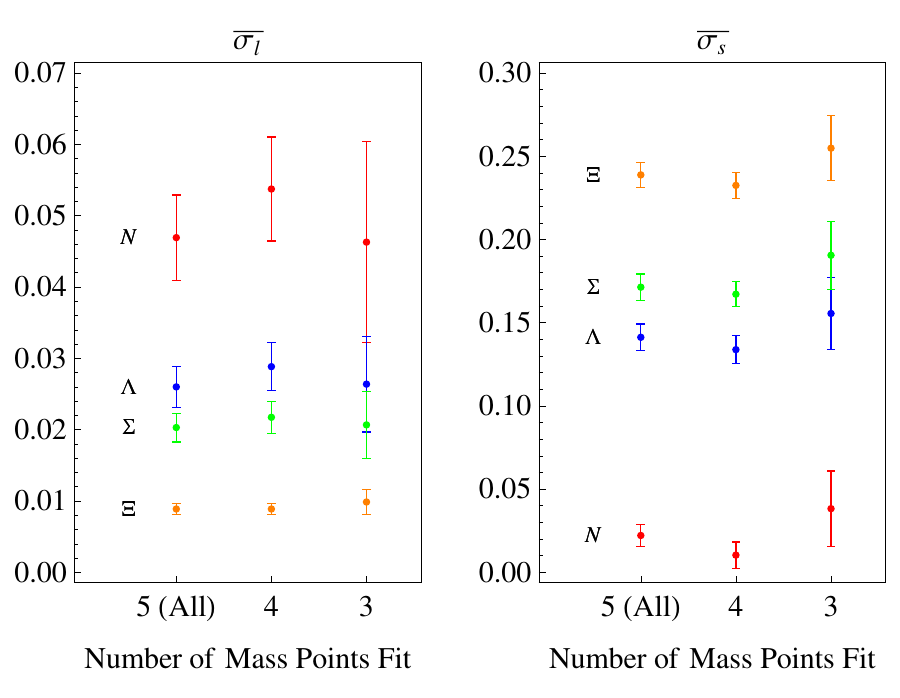}
\caption{Dimensionless baryon sigma terms, evaluated using a dipole
  regulator, based on fits to the NPLQCD results at the lightest 5
  (all), 4, and 3 pseudoscalar mass points.}
\label{fig:varypt}
\ec
\end{figure}

As we fit baryon mass functions to lattice data over a range of
pseudoscalar masses significantly larger than the physical values, it
is prudent to check the consistency of our results as the analysis
moves outside the power-counting regime (PCR), where higher order
terms may become significant.  By performing our fit to progressively
fewer data points, that is, by dropping the heaviest mass points, we test the scheme dependence of our evaluation.
The results are consistent, and largely independent of the truncation
of the data.  This can be seen clearly in Figure~\ref{fig:varypt},
which shows the variation of the dimensionless baryon sigma terms as
progressively fewer data points are used for the fit to the octet
masses. The points shown correspond to an evaluation with a dipole
regulator, and error bars are purely statistical.

\begin{center}
\begin{table}[tbh]
\begin{ruledtabular}
\begin{tabular}{c c c c c}
$B$ & Mass (GeV) & Experimental & $\overline{\sigma}_{Bl}$ & $\overline{\sigma}_{Bs}$ \\ 
\colrule
$N$ & 0.959(24)(9) & 0.939 & 0.047(6)(5)  & 0.022(6)(0) \\
$\Lambda$ & 1.129(15)(6) & 1.116 & 0.026(3)(2) & 0.141(8)(1) \\
$\Sigma$ & 1.188(11)(6) & 1.193 & 0.020(2)(2) & 0.172(8)(1) \\
$\Xi$ & 1.325(6)(2) & 1.318 & 0.0089(7)(4) & 0.239(8)(1) \\
\end{tabular}
\caption{Extracted masses and sigma terms for the physical
  baryons. The first uncertainty quoted is statistical, while the
  second results from the variation of various chiral parameters and
  the form of the UV regulator as described in the text. The
  experimental masses are shown for comparison.}
\label{table:sigmas}
\end{ruledtabular}
\end{table}
\end{center}

\begin{figure}[tb]
\bc
\includegraphics[width=0.48\textwidth]{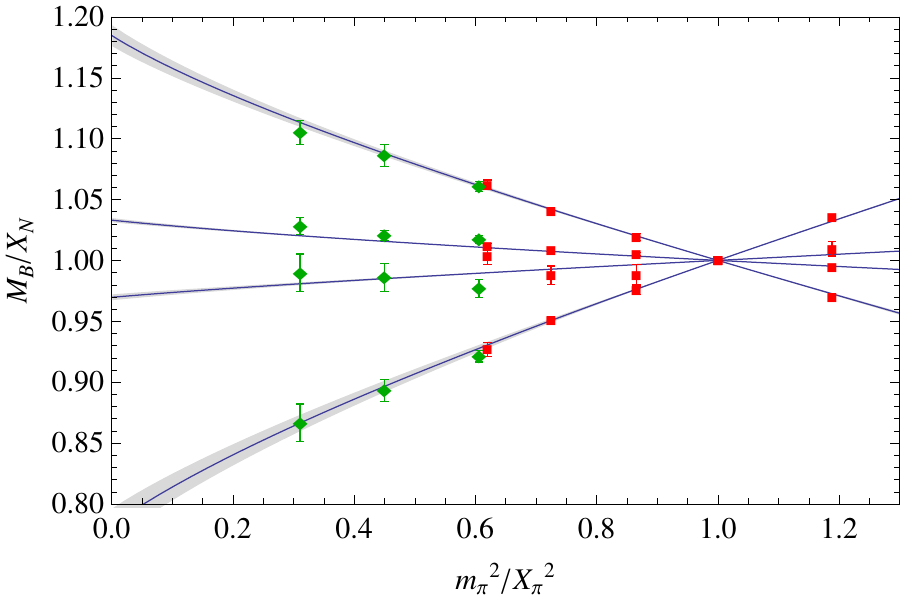}
\caption{Prediction of UKQCD-QCDSF lattice data, based on our fit to the PACS-CS octet baryon mass simulation. Red (square) and green (diamond) points correspond to $24^3$ and $32^3$ lattice volumes respectively. Error bands shown are purely statistical, and incorporate correlated uncertainties between all fit parameters.}
\label{fig:zan}
\ec
\end{figure}
\begin{figure}[tb]
\bc
\includegraphics[width=0.33\textwidth]{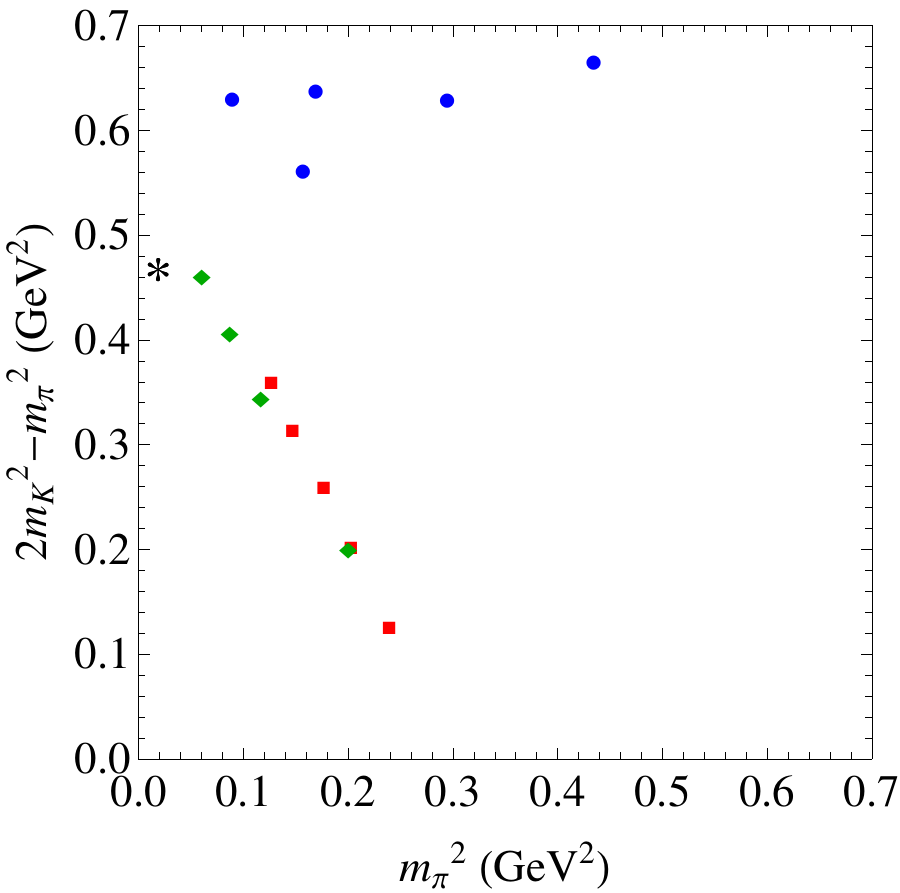}
\caption{Locations of lattice QCD simulations by the PACS-CS Collaboration (blue circles), and UKQCD-QCDSF Collaboration (red and green squares and diamonds) in the $m_l-m_s$ plane. The star denotes the physical point. Figure~\ref{fig:zan} shows the fit to the PACS-CS data only, evaluated at the UKQCD-QCDSF simulation quark masses.}
\label{fig:pointlocs}
\ec
\end{figure}

To further test our claim that the
fitted mass functions accurately describe the variation of the baryon
masses with quark mass, we compare our extrapolation with independent lattice data along a very different trajectory in the $m_l-m_s$ plane, as compared to the fit domain.
Most lattice simulations, including that of the PACS-CS Collaboration,
hold the simulation strange quark mass fixed near the physical value,
and progressively lower the light quark mass to approach the physical
point. However, the UKQCD-QCDSF Collaboration has recently presented an
alternative method of tuning the quark masses, in which the singlet
mass $(2m_K^2+m_\pi^2)$ is held
fixed~\cite{Bietenholz2011}. This procedure constrains the simulation
kaon mass to always be less than the physical value. In comparison,
the traditional trajectory in the $m_\pi-m_K$ plane necessarily keeps
the kaon mass larger than the physical value.

The close match between our fit to the
PACS-CS points and the UKQCD-QCDSF lattice data, shown in Figure~\ref{fig:zan}, is
extremely encouraging. We emphasize that the lines in
Figure~\ref{fig:zan} are \textit{not} a fit to the data shown, but
rather a prediction, resulting from the described fit to the PACS-CS
octet data being evaluated along the UKQCD-QCDSF simulation
trajectory.

All lattice points shown in Figure~\ref{fig:zan} have been shifted, by the procedure described for
the PACS-CS data, to account for finite-volume effects. We chose to use the lattice spacing  $a=0.078$~fm deduced by the UKQCD-QCDSF Collaboration. For
further details of the UKQCD-QCDSF data set, and the normalizations $X_N$,
$X_\pi$, we refer to Ref.~\cite{Bietenholz2011}.

Figure~\ref{fig:pointlocs} illustrates the significance of the prediction shown in Figure~\ref{fig:zan}. While our fit was made only to the PACS-CS data, it successfully reproduces the UKQCD-QCDSF lattice results, at points in the $m_l-m_s$ plane which are substantially different from the PACS-CS simulation trajectory. This very strongly supports our claim that the sigma terms, which correspond to derivatives in the plane shown in Figure~\ref{fig:pointlocs}, are accurately determined by our fit.

To extract the sigma commutators from our baryon mass functions, we use the Feynman-Hellman relation~\cite{Feynman:1939},
\begin{equation}
\sigma_{Bq} = m_q \frac{\del M_B}{\del m_q},
\end{equation}
and, as above, replace quark masses by meson masses squared: $m_l \rightarrow m_\pi^2 /2$ and $m_s \rightarrow (m_K^2-m_\pi^2 /2 )$.
For the case of the nucleon, we recall the alternative conventional notation to quantify the strangeness content, namely the kaon sigma term
\begin{equation}
\sigma_{KN} = \frac{1}{2} (m_l + m_s) \langle N | \overline{l} l + \overline{s} s | N \rangle.
\end{equation}
As in Equation~\ref{eq:dimlesssigma}, overlines indicate corresponding (dimensionless) quantities normalized by the baryon mass.

A direct measure of the magnitude of the strange quark content of the nucleon relative to its light quark content:
\begin{equation}
y=\frac{2\langle N | \overline{s} s | N \rangle}{\langle N | \overline{u}u+\overline{d}d | N \rangle}=\frac{m_l}{m_s} \frac{2 \sigma_s}{\sigma_{\pi N}},
\end{equation}
can be trivially evaluated given the strange and light quark sigma terms. At the physical point, we find $\sigma_{\pi N} = 45 \pm 6$~MeV, $\sigma_{KN}=300\pm40$~MeV and $\sigma_s = 21 \pm 6$~MeV, corresponding to a $y$-value of $0.04 \pm 0.01$ for $m_l/m_s=0.039(6)$~\cite{Nakamura2010}. This analysis also constrains $\sigma_{\pi N} - \sigma_0$ to be $1.64 \pm 0.53$~MeV. The quoted errors include all systematic and model-dependent uncertainties combined in quadrature. Results for the other octet baryons are made explicit in Table~\ref{table:sigmas}.
%
%

An advantage of the method used here is that we can easily evaluate sigma terms from our fit at any pion or kaon mass. The QCDSF Collaboration has recently presented very precise direct calculations of $\sigma_{\pi N}$ and $\sigma_s$ from lattice QCD, at light quark masses somewhat larger than the physical values~\cite{Bali2011}. At $(m_\pi,m_K)$ values of $(281,547)$~MeV, the Collaboration quotes $\sigma_{\pi N}^{\textrm{QCDSF}}=106(11)(3)$~MeV, and $\sigma_s^{\textrm{QCDSF}}=12_{-16}^{+23}$~MeV. This compares very favorably to the results of our fits at these particular pseudoscalar masses, namely $\sigma_{\pi N}=131(11)(5)$~MeV and $\sigma_s=16(5)(1)$~MeV. Once again, the two uncertainties correspond to the described evaluation of systematic and model-dependent uncertainties.
%

The conclusion of our analysis is clear. By developing closed-form functions for baryon mass as a function of quark mass based on a fit to PACS-CS Collaboration lattice data, we were able to determine precise baryonic sigma terms by simple differentiation. This method allows us to achieve small statistical and model-dependent uncertainties. Moreover, we find excellent agreement with recent direct lattice calculations of these values at unphysical pseudoscalar masses. Decidedly our most significant result is a very precise value for the strangeness nucleon sigma term, namely $\sigma_s=21 \pm 6$~MeV at the physical point.

%
\section*{Acknowledgements}
%
%
We acknowledge helpful discussions with J. Zanotti. This work was supported by the University of Adelaide and the Australian
Research Council through the ARC Centre of Excellence for Particle Physics at the Terascale and grants FL0992247 (AWT) and DP110101265 (RDY).

%
%



\begin{thebibliography}{99}
%
\bibitem{Ji:1994av} 
  X.~-D.~Ji,
  Phys.\ Rev.\ Lett.\  {\bf 74}, 1071 (1995)
%
\bibitem{Bottino:1999ei} 
  A.~Bottino {\it et al.},
  Astropart.\ Phys.\  {\bf 13}, 215 (2000)
%
\bibitem{Ellis:2008hf} 
  J.~R.~Ellis, K.~A.~Olive and C.~Savage,
  Phys.\ Rev.\ D {\bf 77}, 065026 (2008)
%
\bibitem{Giedt:2009mr} 
  J.~Giedt, A.~W.~Thomas and R.~D.~Young,
  Phys.\ Rev.\ Lett.\  {\bf 103}, 201802 (2009)
\bibitem{Underwood2012}
S.~J.~Underwood {\it et al.},
[arXiv:1203.1092 [hep-ph]]
%
\bibitem{Hill:2011be} 
  R.~J.~Hill and M.~P.~Solon,
  Phys.\ Lett.\ B {\bf 707}, 539 (2012)
%
\bibitem{Young:2009ps}
R.~D.~Young and A.~W.~Thomas,
Nucl.\ Phys.\ {\bf A844}, 266C (2010)
%
\bibitem{Young:2009zb}
R.~D.~Young, A.~W.~Thomas,
Phys.\ Rev.\  {\bf D81}, 014503 (2010)
%
\bibitem{Ohki2009}
H.~Ohki {\it et al.},
 PoS \ {\bf LAT2009}, 124 (2009)
%
%
\bibitem{Dinter:2012tt}
 S.~Dinter {\it et al.},
[arXiv:1202.1480 [hep-lat]]

\bibitem{Bali:2011ks}
 G.~S.~Bali {\it et al.}  [QCDSF Collaboration],
 Phys.\ Rev.\ {\bf D85}, 054502 (2012)

\bibitem{Dinter:2011zz}
S.~Dinter, V.~Drach and K.~Jansen,
Int.\ J.\ Mod.\ Phys.\ Proc.\ Suppl.\ {\bf E20}, 110 (2011)

\bibitem{Horsley:2011wr}
 R.~Horsley {\it et al.},
Phys.\ Rev.\ {\bf D85}, 034506 (2012)

\bibitem{Durr:2011mp}
S.~Durr {\it et al.},
Phys.\ Rev.\ {\bf D85}, 014509 (2012)

\bibitem{Babich:2010at}
 R.~Babich {\it et al.},
 Phys.\ Rev.\ {\bf D85}, 054510 (2012)

\bibitem{Takeda:2010cw}
 K.~Takeda {\it et al.}  [JLQCD Collaboration],
 Phys.\ Rev.\ {\bf D83}, 114506 (2011)
%
\bibitem{Toussaint2009}
D.~Toussaint, W.~Freeman [MILC Collaboration],
Phys.\ Rev.\ Lett.\ {\bf 103}, 122002 (2009)
%
\bibitem{Ellis2009}
J.~Ellis, K.~A.~Olive, P.~Sandick
New \ J.\ Phys.\ {\bf 11}, 105015 (2009)
%
%
\bibitem{Bali2011}
G.~S.~Bali {\it et al.} [QCDSF Collaboration],
[arXiv:1111.1600 [hep-lat]]
%
%
\bibitem{Shanahan:2011}
P.~E.~Shanahan, A.~W.~Thomas, R.~D.~Young,
Phys.\ Rev.\ Lett.\  {\bf 107}, 092004 (2011)
%
\bibitem{Aoki:2008sm}
S.~Aoki {\it et al.} [PACS-CS Collaboration],
Phys.\ Rev.\  {\bf D79}, 034503 (2009)
%
%
\bibitem{Leinweber:2003dg}
D.~B.~Leinweber, A.~W.~Thomas, R.~D.~Young,
Phys.\ Rev.\ Lett.\  {\bf 92}, 242002 (2004)
%
%
\bibitem{Hall2011}
J.~M.~M.~Hall {\it et al.}
Phys.\ Rev.\ {\bf D84}, 114011 (2011)
%
%
%
\bibitem{Stuckey:1996qr}
R.~E.~Stuckey, M.~C.~Birse,
J.\ Phys.\ {\bf G23}, 29-40 (1997)
%
\bibitem{Donoghue:1998bs}
J.~F.~Donoghue, B.~R.~Holstein, B.~Borasoy,
Phys.\ Rev.\  {\bf D59}, 036002 (1999)
%
\bibitem{Leinweber:1998ej}
D.~B.~Leinweber, D.~-H.~Lu, A.~W.~Thomas,
Phys.\ Rev.\  {\bf D60}, 034014 (1999)
%
\bibitem{Thomas2011}
A.~W.~Thomas, P.~E.~Shanahan, R.~D.~Young,
Few Body Systems (2012),
[DOI:10.1007/s00601-012-0372-8]
%
\bibitem{Beane:2011pc}
 S.~R.~Beane {\it et al.},
Phys.\ Rev.\ {\bf D84}, 014507 (2011)
\bibitem{Geng2011}
Li-sheng~Geng {\it et al.},
Phys.\ Rev.\ {\bf D84}, 074024 (2011)
%
\bibitem{Khan2004}
A.~Ali~Khan {\it et al.} [QCDSF-UKQCD Collaboration],
Nucl.\ Phys.\ {\bf B689}, 175 (2004)
%
\bibitem{Hall2012}
J.~M.~M.~Hall, D.~B.~Leinweber, R.~D.~Young,
[arXiv:1201.6114[hep-lat]]
%
\bibitem{Guichon1983}
P.~A.~M.~Guichon, G.~A.~Miller, A.~W.~Thomas,
Phys.\ Lett.\ {\bf B124}, 109 (1983)
%
\bibitem{Bietenholz2011}
W.~Bietenholz {\it et al.} [QCDFS-UKQCD Collaboration],
[arXiv:1102.5300[hep-lat]]
%
%
\bibitem{Feynman:1939}
R.~P.~Feynman,
Phys.\ Rev.\ {\bf 56}, 340 (1939)
%
\bibitem{Nakamura2010}
K.~Nakamura {\it et al.} [Partice Data Group],
J.\ Phys.\ {\bf G37}, 075021 (2010)
%
\bibitem{Leinweber:2004}
D.~B.~Leinweber,
Phys.\ Rev. \ {\bf D69}, 014005 (2004)



%
\end{thebibliography}
\end{document}